# Detection and Compensation of Linear Attacks in Cyber Physical Systems through System Identification

Soheila Barchinezhad
*School of Electrical and Computer Engineering*
*University of Tehran*
Tehran, Iran
S.Barchinezhad@ut.ac.ir

Mohammad Sayad Haghighi
*School of Electrical and Computer Engineering*
*University of Tehran*
Tehran, Iran
sayad@ut.ac.ir

*Abstract*— Cyber-physical Systems (CPSs) are vasly used in today's cities critical infrastructure. The cyber part of these systems usually has a network component through which cyber attacks can be launched. In this paper, we first design an intrusion detection system (IDS) by indentifying the plant. We assume the initial operation period of the CPS is attack-free and learn the plant model. Then, we compare the expected output found via the identifier with the real one coming through the feedback link. Any difference greater than a threshold is deemed to be an anomaly. To compensate, once the IDS flags a change in the loop, we restart the system identification to find the new transfer function. With the estimation of the new transfer function at hand, a new controller is designed to keep the system stable. To test the idea, we took a DC motor as the plant and employed ARX identifier. Matlab Simulink environment was used to test the proposed intrusion detection and compensation framework. We applied a set of deception attacks to the forward channel in our experiments. The obtained results prove that our detection strategy works well and timely reacts to anomalies. Moreover, they show that the compensation strategy is also effective and keeps the system stable under such attacks.

*Keywords— Intrusion Detection System (IDS), System Identification, Cyber-Physical Systems (CPS), Compensation, ARX.*

## I. Introduction

Cyber-Physical Systems (CPSs) are control systems which are composed of sensors, actuators, control, and network components. They have been employed in many domains including gas pipelines [1], electric power grids [2], medicine [3], water networks [4], driver assistance systems [5, 6] and autonomous electric vehicles [7]. Using computer networks in these systems for communication and interconnection of components makes them vulnerable to cyber threats [8]. Therefore, appropriate detection and possibly compensation techniques are required to counter such cyber attacks. Designing Intrusion Detection Systems (IDS) for CPSs has gained considerable attention [7] due to the consequences of CPSs failure. IDS is a necessary complement to security mechanisms like firewalls. It detects intrusions and provides forensic reports to system administrator or control center to react. IDSs are classified into signature-based and anomaly-based categories. Signature based methods [9] recognize an intrusion based on a previously-known intrusion or attack characteristics. On the other side, anomaly detection methods [10, 11] identify an intrusion by calculating a deviation from normal system behavior.

Traditional IDSs are inadequate for CPSs as they are unaware of the events happening in the physical part of the system. Cyber intrusion detection focuses mainly on investigating the changes in the physical process outputs which cannot be detected by traditional attack detectors or human observer.

This paper is concerned with the detection issue for CPSs against deception attacks. Deception attacks compromise the integrity of packets, and manipulate the behavior of physical process. Some deception attacks are covert and cannot be easily perceived by human beings. They slightly modify the behavior of the system, but their effect cannot be identified by a human observer. This paper tries to model the process or plant normal behavior, create a residual error locally by comparing the model and plant outputs, and compensating the effect of attack upon crossing of the residual error from the detection threshold. Accordingly, an intrusion detection approach is proposed which detects the unordinary changes in the physical process output by using a statistical based threshold. Once the detector detects the presence of an attack, the attack function identification process is triggered. We assume the attacks are of linear type and can be modelled by a transfer function in the frequency/Laplace domain or its discrete time equivalent (z-transform). After characterizing the attack, the final step of the proposed scheme is to design a compensating controller to thwart the negative effects of attack. We switch to this second controller once the attack is detected and identified.

To evaluate the proposed scheme, we conduct a set of simulations under Matlab Simulink. A DC motor is used as the plant of the control system. We simulate packet modification attack by using TrueTime tools [12]. The simulation results, as presented later, demonstrate that the proposed scheme well detect and control deception attacks which cannot be easily identified by a human being. Moreover, the compensation strategy proves effective in the simulated scenarios.

The rest of this paper is organized as follows. The related studies are given in Section II. The proposed detection and identification scheme as well as designing a compensating controller is presented in Section III. In Section IV, the results of simulations on a sample CPS are reported and discussed. The concluding remarks are given in Section V.

## II. RELATED WORKS

Securing CPSs has emerged as an important issue for most governments in last few years. In this context, studies have been conducted to security of CPSs in many aspects. For example, the security challenges of CPSs have been discussed in detail in [13, 14]. The paper [15] consider the security and privacy issues in intelligent transportation systems which is an example of CPS. Cyber-attacks has received increasing attention after the launch of the Stuxnet worm [16]. Cyber-attacks in modern CPSs are divided in three categories [17]: attacks on the control layer, attacks on the communication network, and attacks on the physical process. Communication network is more vulnerable than other components to cyber-attacks. These cyber-attacks can be classified into denial of service (DoS) and deception attacks [18]. Deception attacks mention to the possibility of damaging the integrity of data packets which change the behavior of sensors and actuators [19, 20]. They are also called false data injection attack [21]. Conversely, DoS attacks [22] decreases access to resources in various ways, such as blocking the communication channel.

The general approach of some researches on the analysis of cyber-attacks is investigating the impact of specific attacks on particular systems. For example, in [23, 24] a covert (stealth) deception attack on the communications network in control and monitoring systems have been introduced which is specific deception attacks. In this paper a hidden agent changes the behavior of the physical system while leaving little residue on the main output of the system. Also in [24], the effect of covert attacks on control systems is investigated. In this paper, a covert deception attack is proposed to disturbance the control system operation based on the information collected by another attack. The author claims that the co-operation of the two attacks can covertly and accurately affect the physical behavior of a system. Covert attacks are also mentioned in . The purpose of this paper is to design a deception attack for service degradation that can be stay stealth and has proved that it can compromises the system stability. In that paper, a gain attack as the false signals are injected to make an overshoot on a DC motor system. Two other papers [25, 26] address the vulnerabilities of control systems, in particular modern irrigation channel control systems. In this papers, the deceptive attacks on both the forward and the backward communication channel are investigated. During this deception attack, the attacker sends false data from sensors or controllers. These attacks can be carried out by obtaining secret keys used by the controllers and physical systems or by conquering some of the sensors or controllers.

Securing the CPSs can be done at several levels. First, some preventive techniques are developed that decreases the occurrences of attacks. But by increasing the attacks in these systems, prevention and resistance techniques can be defeated and the development of intrusion detection techniques is a necessity. Intrusion detection, identification, and compensation is one of the most important defense mechanisms against cyber-attacks which a great deal of researches [27, 28] on security of CPSs and IoT [29] have dealt with it. This mechanism is activated after that attacks occur. Most research in this area has focused only on the issue of detection, and a few of them have payed to identification and compensation of intrusions. In [30], an analytical method for intrusion detection in smart grids based on error analysis is presented. The proposed method is based on the state estimation algorithm which minimizes the sum of squares of error. Paper [31] present a specification based IDS for home area network. Normal behavior of the network is defined through selected specifications and deviations from the defined normal behavior can be a sign of some intrusions. In a similar fashion, in [32] authors developed an algorithm that monitors power flow and detects anomalies. This algorithm uses principal component analysis to separate regular and irregular power flow data. Analysis of the information in this subspace determines whether the power system data has been compromised. Another research proposed a model based intrusion detection and impact reduction for an automatic control system [8]. This paper first illustrates the impact of deception attacks on automatic generation control (AGC) and proposes a general framework for use in intelligent attack control. It actually develops a model-based attack detection algorithm and mitigates the attack effects.

There is a large variety of previous work on intrusion detection in CPSs but they have failed to simultaneously detect and identify the nature of the attack and design a new controller based on detection and identification results. Even methods that claim to perform attack identification, actually do the classifying of intrusions, rather than fully identifying the nature of the attack and its effects.

## III. THE PROPOSED APPROACH

When a CPS works with a communication network, data packets are sent between the controller and plant through the network, and may be attacked. The issue under study in this section is how to detect (linear) attacks which are launched over the network and to control the destructive effect of them.

### A. The Proposed Scheme for Intrusion Detection and Compensation

Deception attacks compromise the integrity of data packets on communication channel between the controller and the physical process. In this paper we study some specific deception attacks which cause a steady state error on plant output or make system unstable. The proposed model for intrusion detection and compensation is shown in Fig. 1.

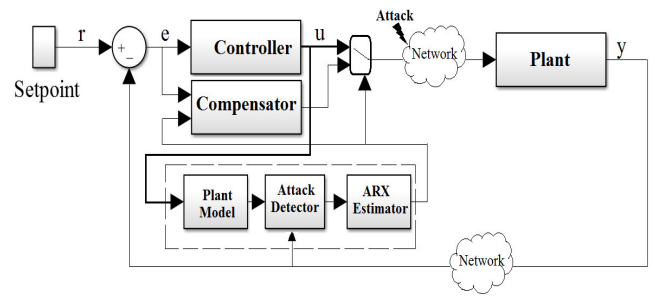

Fig. 1. Block diagram of the proposed intrusion detection and compensation system

It is assumed that the feedback channel is secure and attacks are launched on the forward channel. It is possible to detect the attacked place with the help of additional sensors. It is assumed that the Man in The Middle (MITM) attack [6] is linear and with a transfer function of M(z). The control center tries to detect attacks.

The solution proposed here is placing an identifier at the controller side of the networked system. We could use an observer to identify the plant, however, for the sake of simplicity in presentation here, we assume that the mathematical model of the plant is known to the IDS. If an approximate model of the plant is to be used, we should additionally assume that there is a secure phase at the beginning of system operation, i.e. when the system starts working. During this phase, the identifier learns the normal behavior of the original plant and the information about the residuals are collected by IDS to determine the detection threshold.

Early detection of intrusion is important to allow maintenance or compensation. A common idea is to compare the model output with the system output. If the residual error exceeds a threshold, there must be an attacker in the loop. This of course shall be done when the secure phase is finished. In other words, the difference between the outputs of the model/identifier and the physical plant is calculated and compared against the detection threshold. Both model and the physical plant are fed with the same control signal. Therefore, any significant deviation shows an anomaly. The choice of threshold plays a significant role at the detection step. A common simple thresholding method is η- standard deviation [33]. The residuals which is collected in the secure phase is used to determine the threshold. Imagine the residual error as a random variable with the mean value of $m_R$ and the standard deviation of $s_R$. It can have e.g. normal distribution. In a simple detection strategy, the upper and lower thresholds can be determined as follows:

$$\lambda_{upp} = m_R + \eta s_R \qquad (1)$$
$$\lambda_{low} = m_R - \eta s_R$$

η is a tuning parameter which depends on system sensitivity.

When a linear attack happens in the forward channel, it changes the transfer function of the system. In this case, system identification methods can be used to identify the current dynamic function governing the system (normal system dynamic function plus the attacker function). It will help to design a proper compensator to counteract the effect of the attack. In the proposed scheme, a system identification method (ARX estimator) is used for this purpose. Once the detector sends the detection signal to the estimator, it is activated. When estimation error converges to zero, the dynamics of the new system will be known. Then, the new controller parameters can be calculated for the new system in the way explained in the next section. The controller usually has some design goals, for example, to reduce overshoots or even steady-state error. Design goals can be met provided that the compensating coefficients are set properly. Therefore, reducing the error or increasing the stability can be realized when there is a proper relationship between the detector, identifier and compensator. The default controller has been designed to achieve the performance goals assuming that the system transfer function is the one defined by the legitimate plant. However, in attack scenarios, we need to redesign a substitute controller to achieve the goals using an estimation of the attack function.

## B. System Identificatio

System identification methods can be used to find the dynamics of systems. If the system is linear, linear system identification methods can be used. But for nonlinear systems there are some other identification techniques (such as a neural network)[34].

In linear system identification, dynamic models such as ARX, ARMA, ARMAX, etc. can be employed. The most common model for systems with output feedback is ARX [35] which is demonstrated in Fig.2. ARX is popular because its parameters can be easily estimated by a linear least squares technique.

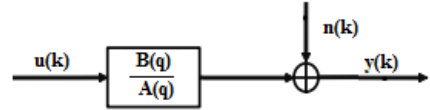

Fig. 2. ARX model [35]

u(k) and y(k) are the discrete system input and output. All disturbances are transferred to the output in the noise n(k). B(q) and A(q) are the numerator and denominator of the transfer function of a system by parameters $a_1, a_2, ..., a_l$ and $b_1, b_2, ..., b_m$. The purpose of identifying the parameters is to obtain values that minimize the error between the model output and the system output. The best choice for merit function is the least sum of squares of error. The ARX model estimator will be as Fig. 3 and Eq. (2).

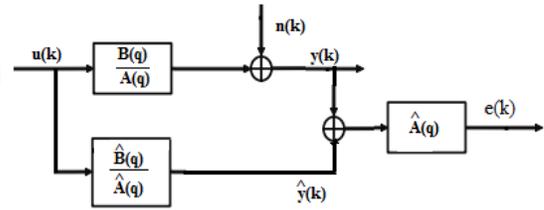

Fig. 3. ARX estimator [35]

In Fig. (3), $\hat{y}(k)$ is the model output for sample $k$.

$$\hat{y}(k|k-1) = B(q)u(k) + (1 - A(q))y(k) \qquad (2)$$

which can be rewritten as

$$\hat{y}(k|k-1) = b_1 u(k-1) + \cdots + b_m u(k-m) \\ -a_1 y(k-1) - \cdots - a_l y(k-l) \qquad (3)$$

assuming deg(A)=l and deg(B)=m. With Eq. (3) the prediction error of an ARX model is,

$$e(k) = A(q)y(k) - B(q)u(k) \quad (4)$$

We can state that
$$y = X\theta \quad (5)$$
with
$$y = \begin{bmatrix} y(m+1) \\ y(m+2) \\ \cdot \\ \cdot \\ \cdot \\ y(N) \end{bmatrix}, \quad \theta = \begin{bmatrix} a_1 \\ a_2 \\ \cdot \\ \cdot \\ a_l \\ b_1 \\ b_2 \\ \cdot \\ \cdot \\ b_m \end{bmatrix}$$

and
$$X = \begin{bmatrix} -y(l) & \cdots & -y(1) & u(m) & \cdots & u(1) \\ -y(l+1) & \cdots & -y(2) & u(m+1) & \cdots & u(2) \\ \cdot & & \cdot & \cdot & & \cdot \\ \cdot & & \cdot & \cdot & & \cdot \\ -y(N-1) & \cdots & -y(N-l) & u(N-1) & \cdots & u(N-m) \end{bmatrix}$$

in which $N$ is the number of available data samples. Then to obtain the least sum of squares of errors, the optimal parameters of the ARX linear model will be as follows:
$$\hat{\theta} = (X^T X)^{-1} X^T y \quad (6)$$

Finally, the calculation of the parameters at the arbitrary step $k$ is done using the parameters obtained in step $k-1$. The use of the forget factor will be useful for tracking the parameters, i.e. reducing the value of the old data compared to that of the new data being entered. This algorithm is capable of tracking the system parameters. By increasing the number of samples to infinity, $\hat{\theta} \to \theta$.

### C. Designing the Controller

The block diagram of a classic well known feedback control system is shown in Fig. 4, where C(z) is the controller function, G(z) represents the system (plant) function, and signals r, e, u and y denote the reference value, control error, control signal and sensed output variable, respectively. In our case, G(z) is actually composed of the attacker's transfer function (M(z)) and the original plant transfer function (H(z)).

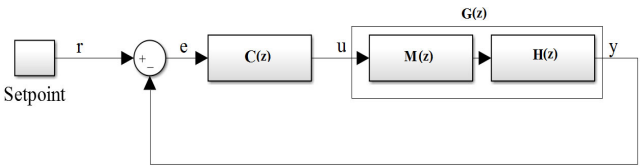

Fig. 4. Block diagram of a feedback control system

There are a veriety of controllers which are used in CPSs and other industrial applications. Proportional, integral and derivative (PID) controller [36] is a kind of controller which is widely used in these systems. It tries to minimize difference between a setpoint and a received sensed or measured variable along the feedback line and accordingly applies a control signal to improve the dynamic response. Generally speaking, the proportional gain reduces error signal, the integral controller reduces steady-state error and the derivative controller improves transient response. The PID controller continues time transfer function can be shown as [37]:
$$C(s) = k_p + \frac{k_i}{s} + k_d s \quad (7)$$

Where $k_p, k_i$ and $k_d$ are the coefficients for the proportional, integral and derivative terms.

The essential problem in designing a PID controller is to determine its parameters which guarantee stabilization of the loop. Many studies have worked on the computation of all stabilizing P, PI and PID controllers parameteres [38, 39]. Consider the single-input single-output control system where,
$$G(z) = \frac{A(z)}{B(z)} \quad (8)$$
is the plant to be controlled.

We take a PID controller for example though a generic PID controller could be designed similarly. C(z) is a descrete PID controller which by approximation, can be written as:
$$C(z) = k_p + \frac{k_i T_s}{z-1} + k_D \frac{N}{1 + NT_s \frac{1}{z-1}} \quad (9)$$
where in implementation, $T_s$ is the sampling time and $N$ is the filter coefficient. The problem is to compute the parameters of PID in Eq. (9) that stabilize the system of Fig. 4. The closed loop transfer function of the system shown in Fig. 4 (by considering the G(z) and C(z) as Eq. (8) and Eq. (9)) is:
$$\frac{Y(z)}{R(z)} = \frac{G(z)C(z)}{1 + G(z)C(z)} \quad (10)$$

The stability of such a system is determined by the location of the poles of the closed loop transfer function in the z-plane or, the roots of the closed loop characteristic polynomial $1 + G(z)C(z) = 0$. When there is no attack, $G(z) = H(z)$.

For the system to be stable, all the roots of carachteristic equation must lie in the unit circle in the z-plane. The system is unstable if any closed loop pole lies outside the unit circle. In our scheme, we want to control the attacked system by a new compensator $C'(z)$. This compensator is a PID controller that replaces the original controller once attack is detected. We assume the identified attacked system by ARX estimator is shown by $G'(z)$, which is presumably an approximation of $M(z)H(z)$. Then, we are going to determine the parameters $k_p, k_i$ and $k_d$, that make the roots of $1 + G'(z)C'(z) = 0$ stay inside the unit circle.

By solving this equation, the stability region for $k_p, k_i$ and $k_d$ is obtained in a 3D space. The PID controller parameters can be then selected from the stable region.

## IV. SIMULATION AND RESULTS

The proposed scheme is implemented in TrueTime environment [12]. One controller and one plant node are considered which communicate through a Controlled Area Network (CAN) bus [40] with each other. The example plant which we have used in this study consists of a second-order plant (DC motor). DC motor is a popular actuator in control systems. We take an example in which the plant is being normally controlled by a PI controller. The PI controller and plant transfer functions are linear time-invariant (LTI) and the set point of this system is defined by a (unit) step function. In the experiment, the discrete time transfer function of the controller and the plant are,

$$H(z) = \frac{9.96 \times 10^{-7} z + 9.92 \times 10^{-7}}{z^2 - 1.988 z + 0.9881} \quad (11)$$

$$C(z) = \frac{(30.2 z - 29.97)}{z - 1} \quad (12)$$

In order to evaluate the performance of the proposed scheme, simulations are conducted. The first implemented attack is a gain attack which makes the plant unstable. In this attack M(z)=160.

In these simulations, the transmission bit rate is set to 250kbps to match J1939 CAN protocol specifications. Also, the secure time span is 5 seconds from the beginning of the simulation, sampling time ($T_s$) is 0.001s and the attack is launched at $t = 5$s. Fig. 5 shows the normal output/behavior of system when there is no attack. In normal operation, the output of plant (motor speed) rises sharply and gradually converges to one when a step like set point with the value of one is applied.

In the mentioned gain-attack case, IDS detects the attack at t=5.003s, merely a few milliseconds after the launch. At this time, the ARX estimator is enabled. The estimation error reaches 7e-15 by t=5.075s. A small error implies that the identification process of system dynamics is completed. The result of identification process in the simulations is,

$$G'(z) = \frac{1.59359996 \times 10^{-4} z - 1.58720003 \times 10^{-4}}{z^2 - 1.98799999 z + 0.988099999} \quad (13)$$

which is close to $M(z).H(z)$ but not identical. $G'(z)$ is an estimated function of system dynamics ($G(z)$). We take a substitute PI controller ($C'(z)$) as an example and determine $k_i$ and $k_p$ that keep the attacked system stable.

$$C'(z) = k_p + \frac{k_i T_s}{z - 1} \quad (14)$$

To find the stability region, we should determine the parameters $k_p$ and $k_i$ that make the roots of $1 + G'(z)C'(z) = 0$ stay inside the unit circle.

$$1 + G'(z)C'(z) = z^3 + (b_1 - 1 + k_p a_1)z^2 \quad (15)$$
$$+ (b_2 - b_1 + k_p a_2 - k_p a_1 + T_s k_i a_1)z$$
$$+ (-b_2 - k_p a_2 + T_s k_i a_2) = 0$$

where $a_1 = 1.59359996 \times 10^{-4}$, $a_2 = -1.58720003 \times 10^{-4}$, $b_1 = -1.98799999$, $b_2 = 0.988099999$ and $T_s = 0.001s$. By using the characteristic polynomial formed on the basis of Eq. (15) and a parametric PI controller defined in Eq. (14), one can find the stabilizing region based on $k_i$ and $k_p$. Once the stability region is found based on the parameters $k_p$ and $k_i$, one can pick a pair that satisfies the desired performance criteria (like small overshoot or steady state error). By solving the Eq.(15), we found the values of parameteres $k_i$ and $k_p$ which make the roots of Eq. (15) stay inside the unit circle. Fig. 6 shows the result. For the mentioned gain attack, selected parameters for compensation are kp=50 and ki=100.

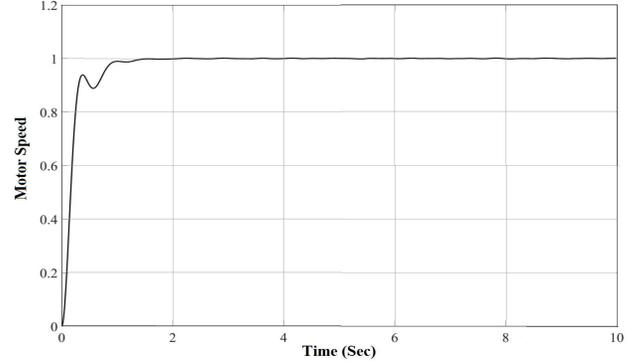

Fig. 5. Behavior (motor speed) of the plant under normal condition.

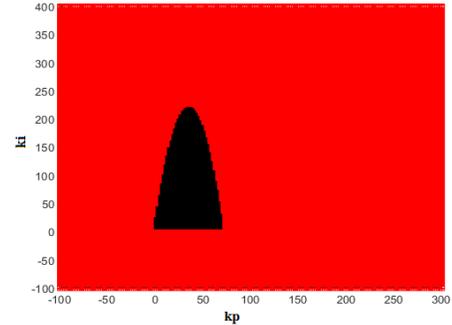

Fig. 6. Stability region in the ($k_p$, $k_i$) plane (for the system with a gain attack=160). The darker area shows where the system is stable.

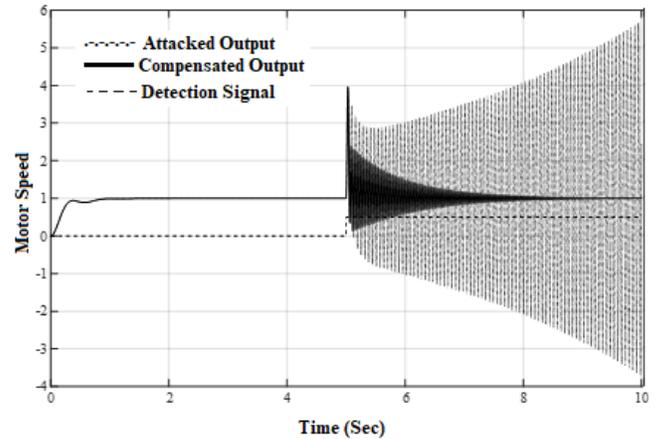

Fig. 7. Behavior of attacked system (for a gain attack=160) with and without IDS and compensator by $k_p$=50 and $k_i$=100.

The output of the case in which attack was launched on the channel, but no IDS or compensator was involved, and also the output of the case in which the proposed intrusion detection, identification and compensation scheme was involved are shown in Fig. 7. Attacker made the system completely unstable. This is obvious from the oscillating output of the system. In Fig.7 IDS flag is raised at 5.003s and the compensation by the substitute controller is activated at 5.075s. It shows the system with IDS and compensator can tolerate the attack more and guarantee the stability. The picked proportional gain and the integral componenet have made system stable and reduced the steady state error of the attacked system, respectively.

The proposed scheme is implemented also with another deception attack which makes a steady-state error. To do so, we use a discrete transfer function (M(z)) as the MITM attacker, similar to the one presented in reference [24],

$$M(z) = \frac{0.7z - 0.7}{z - 1.0001} \quad (16)$$

In this case, attack is detected at t=5.023s, and the estimation error reaches 6e-16 by t=5.12s. The result of ARX estimator for system by mentioned attack is as,

$$G(z) = \frac{6.972 \times 10^{-7} z^2 + 0.028 \times 10^{-7} z - 6.944 \times 10^{-7}}{z^3 - 2.988100049 z^2 + 2.976298898 z - 0.988198859}$$

(17)

For the system with this steady state error attack, stability region based on proportional and integral components of PI compensator has been shown in Fig. 8. Also, Fig. 9 shows the system output. When there is no IDS or compensator, the attacker manages to create a steady state error of magnitude 0.032 which, by given the original controller, does not converge to zero. With the proposed intrusion detection, identification and the designed compensator, the system can tolerate the attacks more. In Fig.9, it is easy to see that the response curve has become better and the steady state error was reduced to 0.005 at t=14s by using PI controller by parameters kp=2000 and ki=1500 as compensator.

## V. CONCLUSIONS

Cyber-physical systems (CPSs) are vulnerable to cyber attacks in their communication layer. Packet manipulation and deception attacks may significantly impair the control system properties such as steady state error, rise and settling time and overshoot/undershoot. In order to recognize the characteristics of the attack and take better decisions, a detection, identification and compensation mechanism is proposed in this paper. In the adversarial model studied, it is assumed that the feedback channel is secure, and attacks are launched on the forward channel. The intrusion detector uses a model of the plant obtained by an identifier at the controller side. The identifier follows the normal behavior of the original plant and can help to detect any deviation from the benign behavior. Once the attack is detected, assuming that it can be approximated by a linear function, an ARX estimator finds the new dynamic function and accordingly, the controller is updated to cope with the attack condition. The goal is to keep the system under attack stable and with desired output. To evaluate the performance of the proposed method, simulations have been conducted using Matlab Simulink and TrueTime tools. A DC motor has been used as the example in simulations. We put the idea to test with a two deception attacks. Simulation results show that the proposed method can effectively help the control system remain stable under such attack scenarios and compensate the negative effects of attacks. For the future work, we aim to extend the proposed approach to support nonlinear systems as well as attacks.

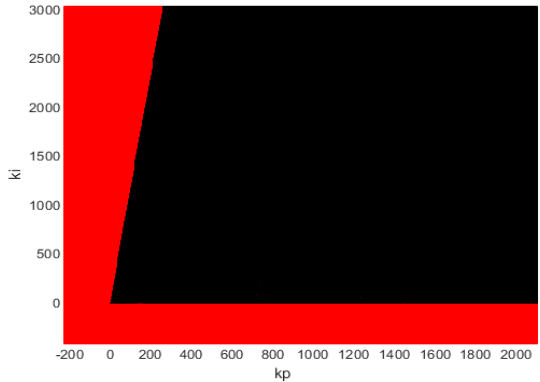

Fig. 8. Stability region in the (k$_p$, k$_i$) plane (for the system with steady state error attack). The darker area shows where the system is stable.

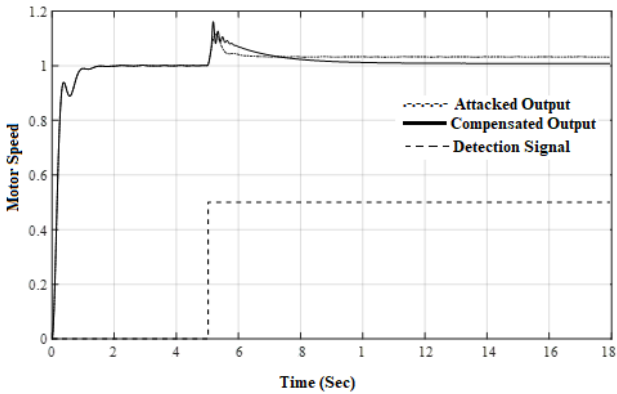

Fig. 9. Behavior of attacked system (with steady state error attack) with and without IDS and compensator by k$_p$=2000 and k$_i$=1500.